\documentstyle[12pt]{article}

\advance\hoffset -0.5cm
\advance\voffset -1.0cm
\newlength{\height}
\divide \height by 3

\begin{document}

\newcommand{\be}{\begin{equation}}
\newcommand{\bea}{\begin{eqnarray}}
\newcommand{\ee}{\end{equation}}
\newcommand{\eea}{\end{eqnarray}}

\def\spur#1{\mathord{\not\mathrel{#1}}} \baselineskip=\height
\begin{titlepage}
\begin{center}
\vskip 0.35in
{{\Large \bf Large-$N$ analysis of $(2+1)$-dimensional Thirring
model}}
\end{center}
\begin{center}
\par \vskip .1in \noindent Deog Ki Hong$^{\dag}$\footnote[1]
{E-mail address: dkhong@hyowon.pusan.ac.kr}
and S. H. Park$^{\ddag}$
\footnote[2]{E-mail address: shpark@phya.snu.ac.kr} \end{center}
\begin{center}
$^{\dag}$Department of Physics, Pusan National University\\
Pusan 609-735, Korea\\
\par \vskip .1in \noindent
$^{\ddag}$Center for Theoretical Physics, Department of Physics\\
Seoul National University, Seoul, Korea
\par \vskip .1in \noindent
\end{center}

\begin{abstract}
We analyze $(2+1)$-dimensional vector-vector type four-Fermi
interaction (Thirring) model in the framework of the $1/N$ expansion.
By solving the Dyson-Schwinger equation in the large-$N$ limit,
we show that in the two-component formalism the fermions
acquire parity-violating mass dynamically in the range of the
dimensionless coupling $\alpha$,
$0 \leq \alpha \leq \alpha_c \equiv {1\over16} {\rm exp}
(- {N \pi^2 \over 16})$.  The symmetry breaking pattern is, however,
in a way to conserve the overall parity of the theory such that
the Chern-Simons term is not induced at any orders in $1/N$.
$\alpha_c$ turns
out to be a non-perturbative UV-fixed point in $1/N$.
The $\beta$ function is calculated to be
$\beta (\alpha) = -2 (\alpha - \alpha_c)$ near the fixed point,
and the UV-fixed point and the $\beta$ function are shown
exact in the $1/N$ expansion.
\vskip 0.2in
\noindent
PACS numbers: 11.15.Pg, 11.15.Ex, 12.20.Ds

\end{abstract}
\end{titlepage}

Recently, there has been a resurgence of interest in four-Fermi
interaction
partly due to the extraordinary heaviness of the top-quark,
compared to other quarks and leptons \cite{topmode}.
One of the key ideas in this
approach is that the 4-Fermi interaction, introduced in the standard
electro-weak theory as a low energy effective interaction, becomes a
relevant operator as the ultraviolet cutoff, $\Lambda$, goes to
$\infty$, due to a strong interaction among fermions.
When the 4-Fermi coupling is
larger than a critical value, the 4-Fermi interaction induces the
condensation of the top-quark, as shown in the original
Nambu-Jona-Lasinio
model \cite{nambu}. Thus the top-quark gets a large mass, and the
electro-weak symmetry breaks dynamically. As described below,
similar dynamical behavior occurs in the $(2+1)D$ Thirring model.

The $(2+1)$-dimensional Thirring model is given in the Euclidean
version by
\be
{\cal L} = i {\overline \psi}_i \spur{\partial} \psi_i +
{\frac{g }{2N}}
\left({\overline \psi}_i \gamma_\mu \psi_i\right)
\left({\overline \psi}_j
\gamma^\mu \psi_j\right),
\label{lagrangian}
\ee
where $\psi_i$ are
two-component spinors and $i,j$ are summed over from 1 to $N$.
The $\gamma$ matrices are defined as
\be
\gamma_3 = \sigma^3,~ \gamma_1 = \sigma^1,~ \gamma_2= \sigma^2,
\label{gammas}
\ee
where $\sigma$'s are the Pauli matrices. Since the four-Fermi
coupling $g$ has a mass-inverse dimension, the model is not
renormalizable in ordinary (weak) coupling expansion.
But, it has been shown renormalizable for $(2+1)$-dimensions
in the large flavor $(N)$ limit \cite{rosen}. It is
therefore sensible to analyze the $3D$ Thirring model in
the large-$N$ expansion.

There are at least two ways of viewing the $3D$ Thirring model in
treating the dimensional coupling constant $g$.
One is taking $g$ as a genuine dimensional parameter
that sets the natural scale of the theory;
for example, the dynamically generated fermions, if any, will be
proportional to this scale; $m_{{\rm dyn}}\sim1/g$.
The another one is to take the dimensional parameter, $1/g$,
as the UV cutoff of the theory;
$g\equiv{\frac{1}{\alpha\Lambda}}$, where $\Lambda$ is
the UV cutoff and
$\alpha$ is a dimensionless coupling. Therefore, in this case,
the only dimensional parameter in the model is the ultraviolet
cutoff. In the continuum limit, the four-Fermi operator
(together with the ultraviolet cutoff),
${\frac{1}{\alpha\Lambda}}\left(\bar\psi\gamma_{\mu}\psi\right)^2$,
becomes a relevant operator in the large-$N$ approximation. In this
approach, if dynamical mass is generated,
it will be independent of the
ultraviolet cutoff, $\Lambda$; it will be the one introduced
in trade of the dimensionless parameter, $\alpha$,
by the so-called dimensional
transmutation, which happens in any renormalizable theories.

The first viewpoint is taken by several authors.
For instance, it has been shown in \cite{kras}
that the $3D$ Thirring model is UV-finite at all orders of $1/N$
since the scale $1/g$ is negligible in the deep UV region.
And also Gomes {\it et. al.} \cite{gom} found in this viewpoint
that the model behaves similarly to ${\rm QED}_{2+1}$ \cite{bow},
which also has a dimensional parameter, $e$, the electric charge;
in both models, the fermion mass is
generated when ${\frac{1}{N}}>{\frac{1}{N_c}}$.
But, as we shall see later,
it is in the second viewpoint that $3D$ Thirring model is
similar to $3D$ Gross-Neveu model \cite{gross}.
Namely, $3D$ Thirring model has a two-phase
structure, parity-broken and parity-unbroken, and the fermion
acquires dynamical mass for strong coupling, $g>g_c$
(or $0\leq\alpha\leq\alpha_c$). Though the model is still UV-finite
perturbatively in $1/N$ expansion, there
exists non-perturbative (in $1/N$) renormalization for $\alpha$. The
coupling is running, $\beta(\alpha)=-2(\alpha-\alpha_c)$, for the
same reason as in the Gross-Neuve model. The UV-fixed point
$\alpha_c$ is found to be ${\frac{1}{16}}\exp(-{\frac{N\pi^2}{16}})$
in the $1/N$ expansion.
The UV-fixed point and the $\beta$ function does not change at all,
even if one includes the higher order corrections, due to
the Ward-Takahashi identity and the UV structure of the theory.
This is in contrast with the result in \cite{kras} presenting
vanishing $\beta$ function. Therefore, we see that two viewpoints
are in many ways different from each other.

Now we start with the effective theory with a UV cutoff.
Introducing an auxiliary field $A_\mu$ to facilitate the
$1/N$ expansion, we can rewrite Eq.~(\ref{lagrangian}) as
\be
{\cal L} = i {\overline \psi}_i \spur{\partial} \psi_i -
       {\frac{1 }{{\sqrt N}}} A_\mu
       \left({\overline \psi}_i \gamma_\mu \psi_i\right) +
       {\frac{1 }{2 }}\alpha \Lambda A_{\mu}^2,
\label{sigmalag}
\ee
where $\alpha \Lambda={\frac{1 }{g}} $.
As was mentioned in \cite{gom}, the theory is consistent for
positive $\alpha$. As we shall see later, for
negative $\alpha$, the theory is unstable showing tachyons
in the four-point fermion Green's function. Eq. (\ref{sigmalag})
is not gauge invariant under the usual gauge transformation on
$\psi$ and $A_\mu$. However, as was claimed in ref. \cite{gom},
Eq. (\ref{sigmalag}) with a gauge fixing term
has a restricted gauge symmetry. In this paper we choose
to work in the Landau gauge.
The Thirring model with $N$ two-component complex spinors has
$U(N)$ global symmetry and parity also. Under $U(N)$,
\be
\psi_i\mapsto \psi^{\prime}_i=g_i^j\psi_j, \; {\rm for }\; g\in U(N)
\ee
and under parity, $P:\; x=(x,y,t)\mapsto x^{\prime}=(-x,y,t)$,
the fermion fields transform as
\be
\psi(x) \mapsto \psi^{\prime}(x^{\prime})=e^{i\delta}\sigma^1\psi(x)
\label{parity}
\ee
One can see that the fermion mass term is parity-odd.
When the number of fermion flavors is even, the model has another
obvious discrete $Z_2$ symmetry, which interchanges half of
the fermions with the other half: $Z_2$
mixes the fermion fields as, for $i=1,\cdots, {\frac{N}{2}}$,
\bea
\psi_i(x) & \mapsto & \psi_{{\frac{N}{2}}+i}(x) \nonumber \\
\psi_{{\frac{N}{2}}+i} & \mapsto & \psi_i(x)
\label{z2}
\eea
We define a new parity $P_4$ which combines the parity for the
two-component spinor with $Z_2$, $P_4\equiv PZ_2$ \cite{parity}.
As described below, in the $(2+1)D$
Thirring model, it is $P$ (not $P_4$) that is spontaneously broken.
The fermion mass is dynamically generated in such a way $P_4$ is
conserved. When $P_4$ is not broken, the Chern-Simons term is
not induced.

Now we will examine the pattern of the spontaneous breaking of
parity. An order parameter for the spontaneous breaking of parity is
the vacuum condensate of the fermion bilinear,
$\left< {\overline \psi} \psi (x) \right>$, which will
be determined once one finds the (asymptotic) behavior of the fermion
propagator \cite{georgi}.

In the $1/N$ expansion one has the following
Dyson-Schwinger gap equation,
\be
-(Z(p)-1)\not\!p+\Sigma(p)=
    {\frac{1}{N}} \int^\Lambda {\frac{d^3k}{(2\pi)^3}}
    D_{\mu\nu}(p-k)\gamma_{\nu}
    {\frac{Z(k)\not\!k-\Sigma(k)}{Z^2(k)k^2+\Sigma^2(k)}}
    \Gamma_{\mu}(k,p-k;p),
\label{gap}
\ee
where $D_{\mu\nu}$ is the photon propagator, $\Sigma$ is
the fermion self
energy, $Z$ is the fermion wave function renormalization constant,
and $\Gamma_{\mu}$ is the vertex function. In the Landau gauge,
the photon propagator is given by
\be
D_{\mu\nu} = {\frac{1 }{p^2}} \left( \delta_{\mu\nu}
     - {\frac{p_\mu p_\nu }{p^2}} \right) \Pi_1 (p^2)
     + \epsilon_{\mu \nu \rho} {\frac{p_\rho }
     {\left|p\right|^2}} \Pi_2 (p^2),
\label{gmunu}
\ee
where $\Pi_1$ and $\Pi_2$ are given by
\be
\Pi_1 (p^2) = {\frac{\alpha \Lambda + \Pi_e }{(\alpha \Lambda +
\Pi_e{})^2/p^2 + \Pi_o^2}}
\label{pione}
\ee
\be
\Pi_2 (p^2) =
   {\frac{\Pi_o }{(\alpha \Lambda + \Pi_e{})^2/p^2 + \Pi_o^2}},
\label{pitwo}
\ee
The resummation technique of the $1/N$ expansion results in
the nontrivial photon propagator as given above. $\Pi_e$ [$\Pi_o$] in
Eq.~(\ref{pione}) [(\ref{pitwo})] is
the even (odd) part of the vacuum polarization which will
be determined once we solve the above coupled Dyson-Schwinger
equations, Eq.'s (\ref{gap}) and (\ref{gmunu}).

Since $Z(p)=1+O(1/N)$ and $\Gamma_{\mu}=\gamma_{\mu}+O(1/N)$,
we may take, at the leading order in $1/N$,
$Z(p)=1$ and $\Gamma_{\mu}=\gamma_{\mu}$ consistently
in Eq.(\ref{gap}). Then, taking trace over the gamma matrix, we get
\be
\Sigma(p)={\frac{1 }{N}} \int^\Lambda\!{\frac{d^3k}{(2\pi)^3}}
{\frac{2\Pi_1(p-k)}{(p-k)^2}} {\frac{\Sigma(k)}{k^2+\Sigma^2(k)}}+
{\frac{1 }{N}}
\int^\Lambda\!{\frac{d^3k}{(2\pi)^3}}
{\frac{(p-k)\cdot k}{\left|p-k\right|^3}}
{\frac{\Pi_2(p-k)}{k^2+\Sigma^2(k)}}
\label{mass}
\ee
The magnitude of dynamically generated mass must be small, compared
to the cutoff, $\Lambda$, of the theory, in the $1/N$ approximation
\cite{bow}. We may therefore assume $\Sigma(p)\ll p \ll \Lambda$.
The vacuum polarization tensor takes then a simple form;
\be
\Pi_e(p) = {\frac{\left| p \right|}{16}}
\label{pie}
\ee
\be
\Pi_o(p) =
    {\frac{1}{N}}\sum_{i=1}^NM_i {\frac{1 }{4 \left|p\right|}},
\label{pio}
\ee
where $M_i\simeq\Sigma_i(0)$, the mass of the $i$-th fermion,
In general, it is hard to find $M_i$ by solving the gap equations
directly. But, following the same
argument of Coleman and Witten \cite{coleman},
one can easily show that the
magnitude of $M_i$ is independent of $i$ in the large
$N$ limit \cite{rajeev}. Therefore, it is reasonable
to assume that $M_i = M$ for $i = 1,...,N-L$
and $M_i = - M$ for $i = N-L+1,...,N$, as is done in \cite{bow}.
For momenta $p$ such that $M \ll p\ll \Lambda$,
\be
{\frac{\Pi_1(p) }{p^2}} = {\frac{1 }{\alpha \Lambda +
{\frac{\vert p \vert}{16}}}}
\label{pion}
\ee
\be
{\frac{\Pi_2(p) }{p^2}} = {\frac{M \theta }{4 \left|p\right|}}
{\frac{1 }{(\alpha \Lambda + {\frac{\vert p \vert}{16}} )^2}},
\label{pitw}
\ee
where $\theta = 1 - {\frac{2L }{N}}$, a parameter characterizing
the parity($P_4$) violation of the theory.
Now we will show that $\theta = 0$ admits a
consistent solution of the gap equation.
Taking the fermion self energy at zero momentum from Eq. (\ref{gap})
and letting $M_i\simeq\Sigma_i(0)$, we find that
\be
M_i = {\frac{2 }{N}} \int^\Lambda{\frac{d^3k }{(2 \pi)^3}}
      {\frac{M_i }{k^2+ M^2}} {\frac{1 }{\alpha \Lambda +
      {\frac{ \vert k \vert}{16}}}} - {\frac{1 }{N}}
      \int^\Lambda{\frac{d^3k }{(2 \pi)^3}}
      {\frac{k^2 }{k^2 + M^2}}{\frac{M\theta }{4 \left| k \right|}}
      {\frac{1 }{(\alpha \Lambda + {\frac{ \vert k\vert}{16}})^2}}.
\label{mas}
\ee
The above equations are consistent only if
$\theta = 0 + O(1/N)$ for $M \neq
0$. Thus $P$ is broken, but $P_4$ is not. One can understand
this result on the basis of Vafa and Witten's argument \cite{vafa}.
The Euclidean fermionic determinant
$Det (\spur{\partial} + M + i\spur{A})$ picks a parity-violating
phase that depends at low momenta on the sign of the fermion mass.
This phase factor increases the ground state energy,
and so the model with the lowest ground state energy should be
the one in which the overall phase is minimized.
Since $Det (\spur{\partial} + M + i\spur{A})$ is the complex
conjugate of $Det (\spur{\partial} - M + i\spur{A})$,
the overall phase vanishes provided the number of positive mass
fermions is equal to the number of negative ones, which is possible
only when $N$ is even.

Actually, one can show further that $\theta =0$ at all orders of
$1/N$ using non-renormalization theorem of
Coleman and Hill \cite{cole}. The requirement
of this non-renormalization theorem is the gauge invariance and
analyticity of the one-loop $n$-photon functions, which is fulfilled
in the $\alpha \leq\alpha_c$ in this model. In this range of
the coupling $\alpha$, the
Chern-Simons term is not subject to the radiative corrections
beyond one loop, thus the one loop result is exact. One speculates,
however, finite radiative correction to the Chern-Simons term in
the other phase due to the lack of analyticity of the fermionic
loops in the infrared region \cite{foot1}.

Then above Vafa and Witten's argument ensures the cancellation of the
Chern-Simons term generated from each fermion and this explains the
absence of the corrections at all.

Following the Cornwall, Jackiw, and Tomboulis
formalism \cite{cornwall}, we calculate the effective potential
of the operator expectation value
$\left< {\overline \psi} \psi (x) \right>$.
At the extrema it is found to be
\be
V = {\frac{N }{2 \pi^2}} \int^\Lambda \! dp p^2 \left[
    {\frac{\Sigma^2 }{p^2+ \Sigma^2}} -
    {\rm ln} \left(1 + {\frac{\Sigma^2 }{p^2}} \right) \right].
\label{eff}
\ee
This is the same expression as in ${\rm QED}_{2+1}$ in the
$1/N$ expansion \cite{bow}. It can be easily seen from
Eq. (\ref{eff}) that any nontrivial
solution has a lower energy than the perturbative vacuum solution,
$\Sigma(p)=0$. Therefore, once such a parity-breaking solution is
found, it is always energetically favored to the symmetric one.
Our solution to the gap equation has thus lower vacuum energy
than the trivial solution.

Rewriting the Eq. (\ref{mas}) when $\theta = 0$,
\be
M = {\frac{2 }{N}} \int^\Lambda \! {\frac{d^3k }{(2 \pi)^3}}
    {\frac{M }{k^2+ M^2}} {\frac{1 }{\alpha \Lambda +
    {\frac{\vert k \vert}{16}}}}.
\label{ma}
\ee
The above equation indicates that there is a two-phase structure.
When $\alpha > \alpha_c$, the parity symmetry is manifest,
where the critical value $\alpha_c$ is defined by the following
equation
\be
1 = {\frac{2 }{N}} \int^\Lambda \! {\frac{d^3k }{(2 \pi)^3}}
    {\frac{1 }{k^2}}{\frac{1 }{\alpha_c \Lambda +
    {\frac{\vert k \vert}{16}}}}.
\label{c}
\ee
If $\alpha \leq \alpha_c$, non-trivial parity-violating fermion
mass is generated in a way to preserve the total parity symmetry
of the theory.

{}From Eq. (\ref{c}), one obtains
\be
\alpha_c = {\frac{1}{16}} {\rm exp} (- {\frac{1 }{16}} N \pi^2).
\label{c1}
\ee
{}From the above equation, one sees the non-perturbative nature in
the phase transition point. The $1/N$ factor in
the gap equation (\ref{ma}) has been
traded to the non-analytic factor $N$ in $\alpha_c$.
The similar phenomena
can be seen in the exponential hierarchy between
the dynamically generated
fermion mass and the cutoff in ${\rm QED}_{2+1}$ \cite{bow}.
The parity violating region, $0 \leq \alpha \leq \alpha_c$ is
very small, so the theory has the parity symmetry for almost
all the positive region of $\alpha$.

The cutoff dependence of the bare coupling is determined
by the requirement that the gap equation, Eq. (18), be
independent of the UV cutoff,
$\Lambda$, as taken to $\infty $. In the vicinity of $\alpha _c$,
we get the $\beta$ function for $\alpha $ given by
\be
\beta (\alpha ) \equiv
     \Lambda {\frac{\partial \alpha }{\partial \Lambda }}
     = - 2 (\alpha - \alpha _c).
\label{beta}
\ee
$\alpha $ increases (decreases) to $\alpha _c$ when
$\alpha <\alpha_c\;(\alpha >\alpha _c)$.
$\alpha _c$ is the UV-fixed point of both phases
of the Thirring model. The above equation shows that $\alpha _c$ is
a UV-fixed point which is again non-perturbative in $1/N$
(it has nonanalytic dependence on $1/N$).
Therefore, it is not feasible to study the theory in
the vicinity of $\alpha_c $ with the perturbative $1/N$ expansion.
To observe the effects of the running of the coupling $\alpha $
as in Eq. (\ref{beta}), one need to go beyond
the perturbative expansion; one need to solve, for example,
the Dyson-Schwinger equation, just as is done in this paper.

As we shall see later,
the theory is UV-finite at any finite orders of $1/N$.
Within the $1/N$ expansion the UV behavior of the theory is
more improved than the one in the Gross-Neveu type model.
In the latter, the resummation
technique of the $1/N$ expansion results in the UV dimension of the
auxiliary field $\sigma =\bar \psi \psi $ being 1 (2 classically).
This is a key point of the renormalizability of the theory.
In the Thirring model the mass of the photon propagator
in Eq.(\ref{gmunu}) should not be neglected at high momenta.
It renders all the integration UV-finite \cite{foot2}.

This can be checked by a direct calculation.
For instance, the fermion wave
function renormalization $Z$ at O($1/N$) is given by Eq.(7).
It is not difficult to see that the integration is finite.
This (perturbative) UV finiteness of the theory is consistent
with the non-perturbative nature of
the UV-fixed point $\alpha _c$.

As we mentioned in the introductory part of
this paper, the mass $M$ is that it is {\it not} a value
which can be determined as in ${\rm QED}_{2+1}$ \cite{bow}
but a parameter as in Gross-Neveu type model \cite{rosen}.
$M$ is a physical quantity, in fact it
is the pole mass of the fermions, and therefore it should be
independent of $\Lambda $. As in the case of Gross-Neveu model,
one may interpret Eq. ({\ref{ma}) as {\it fine-tuning}
the coupling $g$ in order to have $M\ll \Lambda $.
In other words, for $M\ll \Lambda $, the coupling is tuned to be very
close, or equal, to the critical value Eq. (\ref{c1}).

The four-point fermionic Green's function, in leading order, is
given by the photon propagator in Eq. (\ref{gmunu}).
We see that there are no tachyons for $g>0$, so the theory is
consistent in that region. For $g<0$, the Green's function shows the
existence of tachyons.

Finally, we show that the features of the $3D$ Thirring
model we have found are exact in the $1/N$ expansion
even if one includes the higher order
corrections. To go beyond the leading order, one should solve the
Dyson-Schwinger equation keeping the higher order corrections in the
propagators and the vertex function.
But, by following general arguments, the
Dyson-Schwinger equation takes a simpler form. Firstly,
since $\theta =0$ in all orders in $1/N$ as we showed earlier,
we can set $\Pi _o(p)=0$ in the photon propagator. Secondly,
beyond the leading order, the magnitude of $M_i$
is same for all flavors, $\left| M_i\right| =M$, which can be seen
easily by an argument similar to that of
Coleman and Witten \cite{coleman}. Therefore,
the gap equation Eq.(\ref{ma}) becomes
\be
M = {\frac 2N} \int^\Lambda \!{\frac{d^3k}{(2\pi )^3}}
{\frac M {Z(k)k^2+M^2}} {\frac 1{\alpha \Lambda +\Pi _e(k)}}
\Gamma (k,0;k),
\label{ma1}
\ee
where $\Gamma (k,0;k)=
    {\frac 14}g^{\mu \nu }{\rm Tr}\gamma _\mu \Gamma_\nu (k,0;k)$.
Keeping terms up to $O(1/N)$, we find by explicit calculations that
\bea
Z(k) & = & 1-{\frac{16}{3N\pi ^2}}
            \ln \left( {\frac{16\alpha +1}{16\alpha }}\right)
            +O\left( {\frac 1{\Lambda}}\right),
\label{nlo}
\\
\Gamma_\mu (k,0;k) & = & \gamma_\mu  \left( 1-{16\over 3N\pi^2}
            \ln \left( {\frac{16\alpha +1}{16\alpha }}\right)\right)
            +O\left( {\frac 1{\Lambda}}\right) .
\label{nlo1}
\eea
Similarly, the even part of the vacuum polarization is,
up to the terms in $1/N$,
\be
\Pi _e(k)={\frac{\left| k\right| }{16}}
        +const.\times {\frac 1N}\cdot {\frac{k^2}{\alpha \Lambda }}
        \ln {\frac {\Lambda}{M}}+O\left( {\frac 1{\Lambda}}\right) .
\label{nlo2}
\ee
where $const.$ is a pure number.
The Feynman diagrams relevant to the
corrections are shown in Figure 1.

Above results, Eq.'s (\ref{nlo})- (\ref{nlo2}), show
that the next-to-leading corrections are either finite or
suppressed by $1/\Lambda$; the $1/N$ corrections are
UV-finite. By dimensional counting, one can easily show that the
$3D$ Thirring model is in fact UV-finite in all orders in
the $1/N$ expansion \cite{foot3}.

This is the same result as in \cite{kras},
but the reason is quite different. In
our case, because of $\alpha \Lambda $ in the photon propagator,
the loop integrations are UV-finite.

The dangerous terms in deriving the UV-fixed point
$\alpha _c$ will be the terms which are not suppressed
as $\Lambda \to \infty $.
But, such terms do not occur in any orders in $1/N$ because of the
UV finiteness of the theory.
Since the higher order corrections to the vacuum polarization
are suppressed by $1/\Lambda$,
the equation defining $\alpha _c$ becomes now
\be
1 = {\frac 2{N}} \int^\Lambda \!  {\frac{d^3k}{(2\pi )^3}}
{1\over(1-F(\alpha_c)) k^2} {\frac 1{\alpha _c\Lambda +{\frac{|k|}
{16}}}}\cdot\left(1-F(\alpha_c)\right) +O({\frac 1{\Lambda} })
\label{c2}
\ee
where $F(\alpha _c)={\frac{16}{3N\pi ^2}}
  \ln \left( {\frac{16\alpha _c+1}{16\alpha _c}}\right) +O(1/N^2)$.

We see that in the above equation the contributions from the
$1/N$ corrections in Eq.'s (\ref{nlo}) and (\ref{nlo1})
cancel each other exactly.
This is due to the Ward-Takahashi identity of the gauge-invariance
in the $3D$ Thirring model.
The equation defining $\alpha_c$ is therefore same for all orders in
$1/N$, and the UV-fixed point and the $\beta$ function
we found earlier are in fact exact in the $1/N$ expansion.

In summary, we have analyzed
the $(2+1)$-dimensional Thirring model in the $1/N$ expansion.
It has been shown that the fermions acquire parity-violating mass
in the way to conserve
the total parity symmetry of the theory.
The Chern-Simons term is not generated at any finite orders in $1/N$.
We have also shown that the theory has a two-phase structure
and a UV-fixed point as in the Gross-Neveu model,
albeit at non-perturbative order in $1/N$. The UV-fixed point,
$\alpha_c={1\over16}\exp(-{N\pi^2\over16})$,
and the $\beta$ function, $\beta(\alpha)=-2(\alpha-\alpha_c)$, which
we have found by solving the Dyson-Schwinger,
are shown
exact in the $1/N$ expansion.

\vskip .1in
\noindent

{\bf Acknowledgments}
\vskip .1in

This work was supported in part by the
Korea Science and Engineering Foundation through SRC program of
SNU-CTP.
This paper was supported in part by NON DIRECTED RESEARCH FUND,
Korea Research Foundation.
We are grateful to C. Lee and R. Jackiw for keeping us informed
of the papers referred in \cite{parity}.

\pagebreak

\pagebreak

\centerline{Figure Caption}
\vskip 0.2in
\noindent
{\bf Figure 1:} The Feynman diagrams relevant to the $1/N$
corrections (wavy line - photon, solid line - fermion).


\begin{thebibliography}{99}
\bibitem{topmode}  W.A. Bardeen, C.T. Hill, M. Lindner,
     {\it Phys. Rev.} {\bf D 41}, 1647 (1990);
     C.T. Hill, {\it Phys. Lett.} {\bf B 266}, 419 (1991);
     S. Martin, {\it Phys. Rev. } {\bf D 45}, 4283 (1992);
     {\it ibid.} {\bf D 46}, 2197 (1992).

\bibitem{nambu}  Y. Nambu and G. Jona-Lasinio, {\it Phys. Rev.}
    {\bf 122}, 345 (1961); {\it ibid.} {\bf 124}, 246 (1961).

\bibitem{rosen}  B. Rosenstein, B. J. Warr, and S. H. Park,
    {\it Phys. Rev. Lett.} {\bf 62}, 1433 (1989);
    {\it Phys. Rep.} {\bf 205}, 59 (1991) and references therein.

\bibitem{kras}  N. V. Krasnikov and A. B. Kyatkin,
    {\it Mod. Phys. Lett.} {\bf A 6}, 1315 (1991).

\bibitem{gom}  M. Gomes, R. S. Mendes, R. F. Ribeiro,
    and A. J. da Silva, {\it Phys. Rev.} {\bf D43}, 3516 (1991).

\bibitem{bow}  T. Appelquist, M. Bowick, D. Karabali,
   and L. Wijewardhana, {\it Phys. Rev.} {\bf D 33}, 3704 (1986);
   R. Pisarski, {\it Phys. Rev.} {\bf D 29}, 2423 (1984);
   T. Appelquist, M. Bowick, D. Karabali, and L. Wijewardhana,
   {\it Phys. Rev.} {\bf D 33}, 3774 (1986);
   D. K. Hong and S. H. Park, {\it Phys. Rev.}
   {\bf D 47}, 3651 (1993).

\bibitem{gross}  D.J. Gross and A. Neuve,
   {\it Phys. Rev. } {\bf D 10}, 1343 (1974).

\bibitem{parity} R. Jackiw and S. Templeton,
   {\it Phys. Rev.} {\bf D 23}, 2291 (1981);
   S. Deser, R. Jackiw, and S. Templeton,
   {\it Ann. Phys. } {\bf (NY) 140}, 372 (1982);
   {\bf (E) 185}, 406 (1988);
   {\it Phys. Rev. Lett.} {\bf 48}, 975 (1982);
   {\bf (C) 59}, 1981 (1987);
   G.W. Semenoff and L.C.R. Wijewardhana,
   {\it Phys. Rev. Lett.} {\bf 63}, 2633 (1989).



\bibitem{georgi}  A. Cohen and H. Georgi,
  {\it Nucl. Phys.} {\bf B 314}, 7 (1989).

\bibitem{coleman}  S. Coleman and E. Witten,
  {\it Phys. Rev. Lett.} {\bf 45}, 100 (1980).

\bibitem{rajeev}  G. Ferretti, S.G. Rajeev, and Z. Yang,
  {\it Int. J. Mod. Phys.} {\bf A7} (1992) 7989.

\bibitem{vafa}  C. Vafa and E. Witten
  {\it Nucl. Phys.} {\bf B234}, 173 (1984).

\bibitem{cole}  S. Coleman and B. Hill,
 {\it Phys. Lett.} {\bf B159}, 184 (1985);
 T. Lee, {\it Phys. Lett.} {\bf B171}, 247 (1986);
 D. K. Hong, T. Lee, and S. H. Park,
 {\it Phys. Rev.} {\bf D48}, 3918 (1993).

\bibitem{foot1} The absence of the infinite corrections
from the ultraviolet divergences persists in the
$\alpha_c \leq \alpha$ phase.

\bibitem{cornwall}  J. Cornwall, R. Jackiw, and T. Tomboulis,
  {\it Phys. Rev. } {\bf D10}, 2428 (1974).

\bibitem{foot2} In \cite{gom},
the same theory has been analyzed keeping the photon mass
as a parameter, which leads to the similar behavior of $\Sigma(p)$
to the one in ${\rm QED}$.

\bibitem{foot3} Suppose the photon mass in
the propagator is negligible. Then, the most divergent terms are
logarithmic, $\sim\ln\Lambda$, since $3D$ QED is renormalizable
in the $1/N$ expansion.
But, because the photon mass term is proportional to $\Lambda$,
any internal photon line decreases the UV divergence
by dimension one,  and thus the
logarithmic divergence is absent in all orders in $1/N$;
$\int^{\Lambda}{dk\over \Lambda +k}=\ln2$, while
$\int^{\Lambda}{dk\over k}=\ln\Lambda$.


\end{thebibliography}
\end{document}